\documentclass[aps,pra,showpacs,twocolumn]{revtex4}
\usepackage{amsfonts}
\usepackage{amsmath}
\usepackage{amssymb}
\usepackage{graphicx}
\usepackage{epsfig}
\usepackage{color}

\providecommand{\U}[1]{\protect\rule{.1in}{.1in}}

\begin{document}

\title{High order harmonic generation in noble gases using plasmonic field enhancement}
\author{M. F. Ciappina$^{1}$\footnote{Present address: Department of Physics, Auburn University, Auburn, AL 36849, USA}}
\author{T. Shaaran$^{1}$}
\author{M. Lewenstein$^{1,2}$}
\affiliation{$^{1}$ICFO-Institut de Ci\`ences Fot\`oniques, 08860 Castelldefels
(Barcelona), Spain}
\affiliation{$^{2}$ICREA-Instituci\'o Catalana de Recerca i Estudis Avan\c{c}ats, Lluis
Companys 23, 08010 Barcelona, Spain}
\keywords{high-order harmonics generation;strong field approximation;
nanostructures; plasmonics}
\pacs{42.65.Ky,78.67.Bf, 32.80.Rm}

\begin{abstract}
We present theoretical studies of high-order harmonic generation (HHG) in rare gases driven by plasmonic field enhancement. This kind of fields appears when plasmonic nanostructures are illuminated by an intense few-cycle laser and have a particular spatial
dependency, depending on the geometrical shape of the nanostructure. We demonstrate that the strong nonhomogeneous character of the
laser enhanced field plays an important role in the HHG process and significantly extends the harmonic cutoff. Our models are based on numerical solution of the time dependent Schr\"odinger equation (TDSE) and supported by classical and semiclassical calculations.
\end{abstract}

\maketitle
\section{Introduction}

One of the most prominent demonstrations of the nonlinear character of the
interaction of a strong laser with matter is high-order harmonic
generation (HHG)~\cite{keitel,krausz}. In the past two decades, this
phenomenon has been an important subject in both theoretical and
experimental physics, as a reliable tool for generating coherent light
sources in the ultraviolet (UV) to extreme ultraviolet (XUV) spectral range.
HHG based coherent light sources open the possibility for extracting temporal and spatial
information with attosecond and sub-Angstrom resolution~\cite{manfred_rev}.
As a results, HHG has found numerous applications in basic research,
biology, material science, and even lithography~\cite{misharmp}.

The physical mechanism behind HHG can be understood by a semi-classical
three-step model \cite{corkum,sfa}. According to this model, in the first
step, an electron reaches the continuum with no kinetic energy through
tunneling ionization. In the second step, the released electron subsequently
accelerated in an oscillating electric field until it is driven back by the laser field toward its parent ion or
molecule. In the third step, upon its return, it recombines with the core
and emits a high-energy photon, with energy equal to the sum of the
electron kinetic energy and the ionization potential. In this model, the
propagation is treated classically while the ionization and recombination
processes are described quantum mechanically. The maximum photon energy
(cutoff) producible with high harmonic generation is given by $I_{p}+3.17
U_{p}$, where $I_{p}$ is the atomic ionization potential, while $%
U_{p}=I/4\omega^{2}$ is the ponderomotive energy in the laser field of
intensity $I$ and frequency $\omega$.

For high-order harmonic generation to occur, one needs a laser field with intensity
greater than $10^{13}$ W/cm$^{2}$), two orders of magnitude larger
than the output of the current modest femtosecond oscillator. It means an
additional process like chirped-pulse amplification is needed to reach the
required intensity for generation of high harmonics using noble gases. 
Even so, the XUV based on HHG has low duty cycle and efficiency \cite{kim}.

The recent demonstration based on surface plasmon resonances as light
amplifiers could overcome these difficulties~\cite{kim,kling}. By maneuvering
surface plasmon resonances, the laser electric fields can locally be
enhanced by more than 20 dB~\cite{muhl,schuck} without the need of extra
cavities or laser pumping process. Indeed, in such a system the intensity of
the locally enhanced field become large enough to exceed the threshold
intensity for harmonic generation in noble gases. In particular, in the case
of gold bow-tie-shaped nanostructures, the initial laser field (800 nm laser
with intensity $10^{11}$ W/cm$^{2}$) enhanced sufficiently to generate the
7th (114 nm) to the 17st (47 nm) harmonics. Moreover, each nanostructure
could potentially act as a point-like source to generate high harmonic radiation. This
would allow collimation of these coherent radiations by means of constructive
interference. This would open a wide range of possibilities to spatially
rearrange nanostructures to enhance and even shape the HHG spectra~\cite{kim}. Additionally, one
of the most unwanted effects of the harmonic propagation, namely the phase matching,
appears to be not important at all in the HHG using plasmonic fields~\cite{kling}.

The physical mechanism of HHG based on plasmonics can be explained as
follows (for full explanation see ~\cite{kim}). A femtosecond low intensity
laser pulse couples to the plasmon mode and initiates a collective
oscillation among the free charges within the metal. This causes a large
resonant enhancement of the local field inside and at the nanostructure
vicinity. This enhancement is well above the threshold intensity
for generating high harmonics. Consequently, by injecting rare gases into the
site of the enhanced field HHG can be produced. In here, the enhanced field
is spatially inhomogeneous in the region where electron dynamics take place.
In addition, the movement of the electron in the enhanced field is
restricted in space. These two features imply strong modifications in the
harmonic spectra, as was shown by several authors~\cite{husakou,yavuz,ciappi2012,tahirJMO}. 
Furthermore, the influence of the nonhomogeneous character of the laser electric field in the photoelectron 
spectra was recently assessed~\cite{ciappiATI}.

The outcome of Kim et al.~\cite{kim} experiment, however, has been recently
under intense scrutiny~\cite{sivis,Kimreply,corkum_priv}. Between the
points to elucidate are the real intensity enhancement of the input laser
field and the damage threshold of the gold bow-tie nanostructures. According
to the finite-element simulations presented in Ref.~\cite{kim}, the local
field intensity could be enhanced by 4 orders of magnitude (40 dB). It means
for input laser intensities of the order of 10$^{11}$ W cm$^{-2}$ the
enhanced field intensity could reach to the order of 10$^{15}$ W cm$^{-2}$
in the vicinity of the bow-tie nanostructure. On the other hand, from the
measured high-order harmonic spectra it appears that these numbers are
unrealistic. The highest measured harmonic was $n\approx 17$, a limit
corresponds to intensities of the order of $5\times10^{13}$ W cm$^{-2}$ (we
have estimated this value using the well established three step or simple
man's model~\cite{sfa}). It is clear that the measured value for enhanced
field intensity is different from the ones obtained in the finite-element
simulations of Ref.~\cite{kim}. In our theoretical models we have considered
this point using a reduction factor in the field enhancement obtained from
the finite element simulations. Moreover, for this work we use a laser pulse
with a longer wavelength, in order to have larger electron excursion, to
explore the effect of the field non-homogeneity more clearly.

To our best knowledge, apart from Ref.~\cite{kim} there is only one more
experiment in which HHG was produced from noble gas using metallic nanostructures~\cite{kling}. Recently, however, different kinds of nanostructures, e.g.
nanoparticles or nanotips, have been used to generate high energy electrons
directly from the nanostructure in the absence of the gas and to study the
distinct and new characteristics of these nanosytems (see e.g.~\cite{peterprl2006,peterprl2010,peternature,peterjpbreview,ropers})

The paper is organized as follows. In the next section we present a
numerical calculation of HHG by incorporating the actual functional form of
the electric field in the vicinity of a metal bow-tie shaped nanostructure
in the time dependent Schr\"odinger Equation. A classical and semiclassical
analysis of an approximated model for nonhomogeneous fields is presented in
Section 3. We finish with our conclusions and a brief outlook.

\section{Single atom response of plasmonic enhanced fields}

Most of the numerical and semiclassical approaches to study laser-matter
processes in atoms and molecules, in particular high-order harmonic
generation (HHG), are mainly based on the dipole approximation in which the
laser electric field ($\mathbf{E}(\mathbf{r},t)$) and its vector potential
associated ($\mathbf{A}(\mathbf{r},t)$) are spatially homogeneous in the
region where the electron dynamics takes place, i.e. $\mathbf{E}(\mathbf{r}%
,t)=\mathbf{E}(t)$ and $\mathbf{A}(\mathbf{r},t)=\mathbf{A}(t)$~\cite%
{keitel,krausz}. On the other hand, the fields generated using plasmonics
nanosystems, such as metallic nanoparticles, nanotips and gold bow-tie shape nanostructures,
present a strong spatial dependence, which cannot be ignored. From a theoretical viewpoint, the HHG process using
homogeneous fields can be tackled using different approaches and
approximations (for a summary see e.g.~\cite{book1,book2} and references
therein). In this section we employ the Time Dependent Schr\"odinger
Equation (TDSE) in order to study the harmonic radiation generated by a
model atom when it is illuminated by a spatially inhomogeneous electric
field. This in an actual field, which is obtained directly from 3D finite-element (FE) simulations where the real parameters of the bow-tie shaped
nanostructures, as those used in~\cite{kim}, are considered. For linearly
polarized fields, which is the case of our study, the dynamics of an atomic
electron is mainly along the direction of the laser electric field and as a
result it is a good approximation to employ the Schr\"odinger equation in
one spatial dimension (1D-TDSE)~\cite{keitel} which can be written as
follows:
\begin{eqnarray}  \label{tdse}
\mathrm{i} \frac{\partial \Psi(x,t)}{\partial t}&=&\mathcal{H}(t)\Psi(x,t) \\
&=&\left[-\frac{1}{2}\frac{\partial^{2}}{\partial x^{2}}+V_{A}(x)+V_{L}(x,t)%
\right]\Psi(x,t)  \nonumber
\end{eqnarray}
where $V_{A} (x)$ is the atomic potential and $V_{L}(x,t)$ represents the
potential due to the laser electric field. In here, we use for $V_{atom}$
the "quasi-Coulomb" or "soft-core" potential
\begin{eqnarray}  \label{atom}
V_{A}(x)&=&-\frac{1}{\sqrt{x^2+\xi^2}}
\end{eqnarray}
which first was introduced in~\cite{eberly} and since then has been widely used in the
reduced dimensions studies of laser-matter processes in atoms. The required
ionization potential can be defined by varying the parameter $\xi$ in Eq. (\ref%
{atom}). The laser potential $V_{L}(x,t)$ of the laser electric field $E(x,t)$ is given by
\begin{eqnarray}  \label{vlaser}
V_{L}(x,t)&=&E(x,t)\,x.
\end{eqnarray}
In Eq. (\ref{vlaser}), the spatial dependency of $E(x,t)$ can be defined in
terms of a perturbation to the dipole approximation and it reads
\begin{equation}  \label{electric}
E(x,t)=E_0\,f(t)\, (1+h(x))\,\sin\omega t,
\end{equation}
which is linearly polarized along the $x$-axis. In Eq. (\ref{electric}), $%
E_0 $, $\omega$ and $f(t)$ are the peak amplitude, the frequency of the
coherent electromagnetic radiation and the pulse envelope, respectively. In
addition, $h(x)$ represents the functional form of the nonhomogeneous
electric field and it can be written as a power series of the form $%
h(x)=\sum_{i=1}^{N}b_i x^{i}$. The coefficients $b_i$ are obtained by
fitting the actual electric field that results from a FE simulation
considering the real geometry of different nanostructures (See panel (b) of
Fig. 1). In this work we use for the short laser pulse a trapezoidal
envelope given by
\begin{equation}  \label{ft}
f(t) = \left\{
\begin{array}{ll}
\frac{t}{t_{1}} & \quad \text{for $0 \leq t < t_1$} \\
1 & \quad \text{for $t_1 \leq t \leq t_2$} \\
-\frac{(t-t_3)}{(t_{3}-t_{2})} & \quad \text{for $t_2 < t \leq t_3$} \\
0 & \quad \text{elsewhere} \\
\end{array}
\right.
\end{equation}
where $t_1=2\pi n_{on}/\omega$, $t_2=t_1+2\pi n_{p}/\omega$, and $%
t_3=t_2+2\pi n_{off}/\omega$. $n_{on}$, $n_p$ and $n_{off}$ are the number
of cycles of turn on, plateau and turn off, respectively.

We employ $\xi=1.18$ in Eq. (\ref{atom}) such that the binding energy of the
ground state of the 1D Hamiltonian coincides with the (negative) ionization
potential of Ar, i.e. $\mathcal{E}_{GS} = -15.7596$ eV ($-0.58$ a.u.).
Furthermore we assume that the noble gas atom is in its initial state
(ground state (GS)) before we turning the laser ($t=-\infty$) on. Equation (%
\ref{tdse}) is solved numerically by using the Crank-Nicolson scheme~\cite%
{keitel}. In addition, to avoid spurious reflections from the spatial
boundaries, at each time step, the electron wave function is multiplied by a
mask function~\cite{mask}.

The harmonic yield of an atom or molecule is proportional, in the single
active electron approximation, to the Fourier transform of the acceleration $%
a(t)$ of the active electron~\cite{schafer}. That is,
\begin{equation}
D(\omega)=\left| \frac{1}{\tau}\frac{1}{\omega^2}\int_{-\infty}^{\infty}%
\mathrm{d} t\mathrm{e}^{-\mathrm{i} \omega t}a(t)\right|^2
\end{equation}
with $a(t)$ is obtained by using the following commutator relation
\begin{equation}  \label{accel1D}
a(t)=\frac{\mathrm{d}^{2}\langle x \rangle}{\mathrm{d} t^2}=-\langle \Psi(t)
| \left[ \mathcal{H}(t),\left[ \mathcal{H}(t),x\right]\right] | \Psi(t)
\rangle.
\end{equation}
In here, $\mathcal{H}(t)$ and $\Psi(x,t)$ are the Hamiltonian and the
electron wave function defined in Eq. (\ref{tdse}), respectively. The
function $D(\omega)$ is called the dipole spectrum, which gives the spectral
profile measured in HHG experiments. For solving Eq. (\ref{tdse}), the gap
size $g$ of the gold bow-tie nanostructure is taken into account restricting
the spatial grid size (see Figure 1).

\begin{figure}[htb]
\centering
\includegraphics[width=0.4\textwidth]{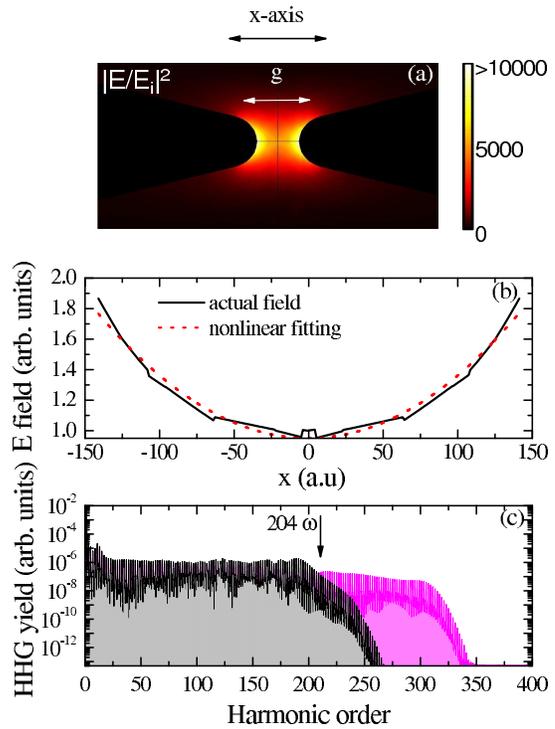}
\caption{(Color online) Panel (a) electric field enhancement in the gap
region of the gold bow-tie nanostructure extracted from the FE simulations;
panel (b) functional form of the electric field $E(x,t)$ where the solid
lines is the raw data obtained from the FE simulations and the red dotted
line is a nonlinear fitting; panel (c) high-order harmonic generation (HHG)
spectra for Ar with ionization potential $\mathcal{E}_{GS}=-0.58$ a.u.,
laser wavelength $\protect\lambda=1800$ nm and intensity $%
I=1.25\times10^{14} $ W$\cdot$cm$^{-2}$ at the center of the gap $x=0$. We
use a trapezoidal shaped pulse, Eq. (\protect\ref{ft}), with $n_{on}=3$, $%
n_{off}=3$ and $n_{p}=4$ (about 60 fs). The gold bow-tie nanostructure has a
gap $g=15$ nm (283 a.u.). Black line indicates the homogeneous case while
magenta line indicates the nonhomogeneous case. The arrow indicates the
cutoff predicted by the semiclassical model for the homogeneous case~%
\protect\cite{sfa}. }
\label{fig:figure1}
\end{figure}

The electric field intensity distribution inside the gap of gold bow-tie
shaped nanoantennas was computed numerically by 3D Finite Element Method
(COMSOL Multiphysics)~\cite{srdjan1}, using the gold optical properties
taken from Ref.~\cite{christy} . The nanoantenna is formed by two identical
(isosceles) triangular gold pads (longest altitude of 600 nm and the
smallest acute angle of 30$^{\circ}$) separated by an air gap $g$. The
apices at corners were rounded (10 nm radius of curvature) to account for
limitation of current fabrication techniques and avoid nonphysical fields
enhancement due to tip-effect. The out of plane thickness is set to 25 nm.
These parameters yield to a dipolar bonding resonance centered at around $%
\lambda=1800$ nm when considering gaps ranging between 12 nm and 15 nm. This
particular value of $\lambda$ was chosen according to the availability of
laser sources~\cite{jens1}. On the other hand, the selected laser wavelength
allows the electron to have excursions of the order of the gap $g$ and
consequently to \textit{confine} its movement. Classically the electron
excursion in an oscillating electric field is given by the so-called quiver
radius $\alpha_0$, which is $\propto \sqrt{I}\lambda^2$ where $I$ is the
laser intensity. For instance, in here, for intensities $I$ of $\sim 10^{14}$
W cm$^{-2}$, $\alpha_0$ can have a value about $\pm 80$ a.u. ($\pm$4.5 nm).

The panel (a) of Fig. 1 displays the calculated electric field enhancement
in the gap of the bow-tie structures when illuminated by a linearly
polarized ($x$-axis) plane wave at 1800 nm. The field-enhancement profile is
extracted for the bow-tie long axis through the middle of the gap, so the
successive problem is reduced to 1D as is shown in panel (b) of Fig. 1.
Additionally we normalize the electric field by setting $E(0,t)=1$. We
observe a typical amplification between 30-40 dB, i. e. 3-4 orders of
magnitude between the input intensity and the intensity at center of the
gap. In a real experiment, however, the enhancement can be smaller than our
calculations; nevertheless, for a similar system, it was shown that one can
obtain values of more than 20 dB~\cite{kim,kling}.

Panel (c) of Figure 1 depicts the harmonic spectra for a bow-tie shaped
nanostructure with gap $g=15$ nm and for a laser wavelength of $\lambda=1800$
nm considering an homogeneous electric field, i.e $E(x,t)=E(t)$ and a
nonhomogeneous electric field using Eq. (\ref{electric}). The enhanced laser
intensity is $I=1.25\times 10^{14}$ W cm$^{-2}$ at the center of the spot ($%
x=0$). In order to reach this intensity value, we consider plasmonic field
enhancements between 25 and 35 dB with input intensities in the range of $%
4\times10^{11}$-$4\times10^{10}$ W cm$^{-2}$. These intensities would be
well below the damage threshold of the nanostructure employed (see e.g.~\cite%
{kim}). In both cases, we use a trapezoidal shaped pulse with three optical
cycles turn on ($n_{on}=3$) and turn off ($n_{off}=3$) and a plateau with 4
optical cycles ($n_{p}=4$), i.e. 10 optical cycles in total which is about
60 fs.

For the homogeneous case we have an harmonic cutoff at around 204$\omega$ as
shown by the arrow in panel (c) of Fig. 1. In fact, our calculation are in
excellent agreement with the semiclassical model~\cite{sfa}. For the
nonhomogeneous case, however, we observe a substantial increase in the
harmonic cutoff, which is about 50 \% higher than the cutoff generated by a
homogeneous electric field. This new feature emerges due to the combination
of the nonhomogeneous character of the electric field and the confinement of
the electron motion~\cite{ciappi2012}. In fact, the laser ionized electron could be
absorbed by the metallic surface (for details see ~\cite{husakou}) and in this
way only the short electron trajectories would be allowed. This circumstance
is comparable to the restriction in the electron motion incorporated in our theoretical 
model~\cite{ciappipra}.  

\section{Classical and semiclassical approaches}

In this section we present classical and semiclassical studies of a
particular case of nonhomogeneous field. In particular, we only consider the
linear term of the series for $h(x)$ (see previous Section). This field
represents, for instance, a first order approximation of the near-field in
the surrounding of a metal nanoparticle~\cite{kling_spie,klingprb}. The main
advantages of this approximation is the possibility to perform a quasi
analytical approach using the Strong Field Approximation (SFA)~\cite{sfa}.
In this way, the main features behind the quantum trajectories are
elucidated and characterized.

\subsection{Strong Field and the Newton equation}

We use Eq (\ref{vlaser}) in order to calculate the Newton equation of motion
of an electron in a nonhomogeneous electric field. In particular, the Newton
equation of motion for an electron in this field is given by
\begin{equation}
\ddot{x}(t)=-\nabla _{x}V_{L}=-x\nabla _{x}E_{eff}(t,x)-E_{eff}(t,x),
\label{Newtonho}
\end{equation}%
where $E_{eff}(t,x)$ gives the effective electric field along the electron
trajectory. For the homogeneous case, in which the laser field does not have
spatially dependency and it is just $E(t)$, the Newton equation reads $\ddot{%
x}(t)=-E(t)$. On the other hand, if the spatial dependence of the enhanced
laser electric field is perturbative and linear with respect to position,
then the effective field can be approximated as
\begin{equation}
E_{eff}(t,x)\simeq E(t)(1+\alpha x),  \label{nhmfield}
\end{equation}
where $\alpha \ll 1$ is a parameter that characterize the strength of the
inhomogeneity.

By substituting (\ref{nhmfield}) into (\ref{Newtonho}), we have
\begin{equation}
\ddot{x}(t)=-E(t)(1+2\alpha x(t))=-E(t,x).  \label{Newtonh1}
\end{equation}

Classically, the electron trajectory can be found by solving Eq (\ref%
{Newtonh1}). In here, we solve this equation by applying the Picard
iteration \cite{diffeq} method and restrict ourselves to the first order
(for more details see \cite{ciappi2012}). In the SFA, the action is defined
in terms of the vector potential field $A(t)$, the counterpart of the
velocity $\dot{x}(t)$ in the classical equation, which corresponds to the
integration of $\ddot{x}(t)$ with respect to $t$. By assuming that the
electron starts its movement from the origin with zero velocity, i.e. $%
x(0)=0 $ and $\dot{x}(0)=0$ , we will have
\begin{equation}
x(t)=\lambda (t)-\lambda (t_{0})-A(t_{0})(t-t_{0}).
\end{equation}
with $\lambda (t)=\int_{0}^{t}dt^{\prime }A(t^{\prime })$. Furthermore, if
we assume the potential field is zero at time $t_{0}$, then
\begin{equation}
x(t)=\int^{t}dt^{\prime }A(t^{\prime }).  \label{newposition}
\end{equation}
By using Eqs. (\ref{newposition}) and (\ref{Newtonh1}), the effective vector
potential along the electron trajectory $A_{eff}(t,x)$ reads
\begin{equation}
A_{eff}(t)=A(t)+2\alpha A_{a}(t),  \label{newpotentialfield}
\end{equation}
where
\begin{equation}
A_{a}(t)=\int^{t}dt^{\prime \prime }A(t^{\prime \prime })-\int^{t}dt^{\prime
\prime }A^{2}(t^{\prime \prime }).
\end{equation}

\subsection{Transition Amplitude}

The strong field approximation (SFA) is based on the assumptions that the
influence of the laser field is neglected when the electrons are bound to
their target atoms and the binding ionic potential is neglected when the
electrons are in the continuum. As a result, the free electrons in the
continuum are described by field-dressed plane waves, which are known as
Volkov states \cite{Gordon1926,Volkov1935}.

Based on the well established model of Lewenstein~\cite{sfa}, the SFA
transition amplitude for HHG reads (in atomic units)
\begin{eqnarray}
b_{\Omega } &=&i\int_{-\infty }^{\infty }\hspace*{-0.2cm}dt\hspace*{-0.1cm}%
\int_{-\infty }^{t}\hspace*{-0.3cm}dt^{\prime }\hspace*{-0.1cm}\int
d^{3}kd_{rec}^{\ast }(\widetilde{\mathbf{k}}(t))d_{ion}(\widetilde{\mathbf{k}%
}(t^{\prime }))  \nonumber \\
&&e^{-iS(\Omega ,\mathbf{k},t,t^{\prime })}+c.c.  \label{Mp}
\end{eqnarray}
with the action
\begin{eqnarray}  \label{newaction}
S(\Omega ,\mathbf{k},t,t^{\prime }) &=&\int_{t^{^{\prime }}}^{t}\hspace{%
-0.1cm}\frac{[\mathbf{k}+\mathbf{A}(\tau )]^{2}}{2}d\tau
+I_{p}(t-t^{^{\prime }})-\Omega t+  \nonumber \\
&&2\alpha \int_{t^{^{\prime }}}^{t}\hspace{-0.1cm}A_{a}(\tau )[\mathbf{k}+%
\mathbf{A}(\tau )]d\tau + \\
&& 2\alpha ^{2}\int_{t^{^{\prime }}}^{t}\hspace{-0.1cm}A_{a}^{2}(\tau )d\tau
\nonumber
\end{eqnarray}
and the prefactors
\begin{equation}
d_{ion}(\widetilde{\mathbf{k}}(t^{\prime }))=\langle \mathbf{\tilde{k}}%
(t^{\prime })|H_{int}(t^{\prime })|\phi _{0}\rangle  \label{prefion}
\end{equation}%
\begin{equation}
d_{rec}(\widetilde{\mathbf{k}}(t))=\langle \mathbf{\tilde{k}}%
(t)|O_{dip}.e_{x}|\phi _{0}\rangle .  \label{prefrec}
\end{equation}
In here, $I_{p}$, $k$, $\Omega $ , $\ H_{int}(t^{\prime })$, $O_{dip}$ and $%
e_{x}$ denote the ionization potential of the field-free bound state $%
\left\vert \phi _{0}\right\rangle $, the drift momentum of the electron in
the continuum, the harmonic frequency, the interaction of the system with
the laser field, the dipole operator and the laser polarization vector,
respectively.

Eq. (\ref{Mp}) describes a physical process in which an electron, initially
in a bound state $|\phi _{0}\rangle $ with energy $I_{p}$, interacts with
the laser field by $H_{int}(t^{\prime })$ at the time $t^{\prime }$ and
ionized into a Volkov state $|\mathbf{\tilde{k}}(t)\rangle $. Subsequently,
it propagates in the continuum from time $t^{\prime }$ to $t$. At the time $t$ 
it is driven back to the parents ion and recombines with the core
emitting high-harmonic radiation of frequency $\Omega $. The SFA is not gauge invariant, 
thus, Eqs (\ref{prefion}) and (\ref{prefrec}) will have different
form in the length and velocity gauges. For the length gauge, the
interaction Hamiltonian and the Volkov wave function are given by $%
H_{int}^{l}(t)=r.E(t^{\prime },x)$ and \ $\widetilde{\mathbf{k}}(t)=\mathbf{k%
}+A_{eff}(t)$, respectively, while for the velocity gauge they are $%
H_{int}^{l}(t)=[k+A_{eff}(t^{\prime })]/2$ and $\widetilde{\mathbf{k}}(t)=%
\mathbf{k}$, respectively. In this work, we consider the length gauge and
assume that the electric field is linearly polarized along $x$-axis.
Furthermore, we consider an hydrogenic $1s$ state for the field-free bound
state $\left\vert \phi _{0}\right\rangle$.

As a result, Eqs. (\ref{prefion}) and (\ref{prefrec}) yield
\begin{equation}
d_{ion}(\widetilde{\mathbf{k}}(t^{\prime }))\propto \frac{\widetilde{\mathbf{%
k}}(t^{\prime })_{x}}{(\widetilde{\mathbf{k}}(t^{\prime })^{2}+2I_{p})^{3}}%
E(t^{\prime }),  \label{newprefion}
\end{equation}
\begin{equation}
d_{rec}(\widetilde{\mathbf{k}}(t))\propto \frac{\widetilde{\mathbf{k}}(t)_{x}%
}{(\widetilde{\mathbf{k}}(t)^{2}+I_{p})^{3}}  \label{newprefrec}
\end{equation}
respectively.

\subsection{Saddle-point equations}

The transition amplitude (\ref{Mp}) was computed employing the saddle-point
method~\cite{Salieres2001}. For this approach one needs to obtain the saddle
points for which the action (\ref{newaction}) is stationary, i.e. for which $%
\partial _{t}S(\Omega ,\mathbf{k},t,t^{\prime })=\partial _{t^{\prime
}}S(\Omega ,\mathbf{k},t,t^{\prime })=\partial _{k}S(\Omega ,\mathbf{k}%
,t,t^{\prime })=0$. The solutions of the saddle point equations are directly
related to the classical trajectories, which allows to demonstrate the
cutoff of the HHG. The stationary conditions upon $t,t^{\prime }$ and $k$
lead to the saddle-point equations
\begin{equation}
\left[ \mathbf{k}+\mathbf{A}(t^{\prime })\right] ^{2}-2\epsilon \beta
(t^{\prime })=-2I_{p}  \label{saddle1}
\end{equation}%
\begin{equation}
\int_{t^{\prime }}^{t}d\tau \lbrack \mathbf{k}+\mathbf{A}(\tau )]+2\alpha
\eta \mathbf{(\tau )}=0  \label{saddle2}
\end{equation}%
and
\begin{equation}
\Omega =\frac{\left[ \mathbf{k}+\mathbf{A}(t)\right] ^{2}}{2}+I_{p}+2\alpha
\beta (t)  \label{saddle3}
\end{equation}%
with
\begin{equation}
\beta (t)=A_{a}(t)[\mathbf{k}+\mathbf{A}(t)]+\alpha A_{a}^{2}(t),
\label{saddle4}
\end{equation}%
and
\begin{equation}
\eta (\tau )=\int_{t^{^{\prime }}}^{t}\hspace{-0.1cm}A_{a}(\tau )d\tau .
\label{saddle5}
\end{equation}%
Eq. (\ref{saddle1}) gives the conservation law of energy for the electron
tunnel ionized at the time $t^{\prime }$. Eq. (\ref{saddle2}) constrains the
intermediate momentum of the electron as well as guarantees that the
electron returns to its parent ion. Finally, (Eq. \ref{saddle3}) expresses
the energy conservation of the electron at the recombination time $t$
and generation of a high frequency photon $\Omega$.

The terms $\beta (t^{\prime })$, $\eta (t)$ and $\beta (t)$ in the Eqs. (\ref%
{saddle1}), (\ref{saddle2}), and (\ref{saddle3}), respectively, give the
nonhomogeneous character of the laser field and they are zero for the
homogeneous case ($\epsilon =0)$. \ For small \ $\epsilon $, the solutions
of the saddle equations are generally complex since Eq. (\ref{saddle1}) has
no real solutions, unless $I_{p}\rightarrow 0$. \ For nonhomogeneous field, $%
3.17U_{p}$, where $U_{p}=E_{0}^{2}/(4\omega ^{2})$ is the ponderomotive
energy, does not directly corresponds to the maximum kinetic energy that the
electron gains in the continuum . Depends on the nonhomogeneous character of
the field, for the positive and negative $A_{a}$ this value is larger and
smaller than $3.17U_{p}$, respectively. 

We now examine the drift momentum $\mathbf{k}$ of the electron at the time
of the tunneling . For that, we take the limit in which the electron reaches
the continuum with zero kinematical momentum, $I_{p}\rightarrow 0$. As a
result, Eq. (\ref{saddle1}) reads
\begin{equation}
\mathbf{k}=-A(t^{\prime })+\epsilon A_{a}(t^{\prime })(1\mp \sqrt{3})
\label{driftmomentum}
\end{equation}%
In here, $\mathbf{k}$ has two different solutions with one lowering and
the other exceeding and the homogeneous drift momenta. This result is
different from the homogeneous case, in which $\mathbf{k}=-A(t^{\prime })$.
This deviation emerges from the strength of the inhomogeneity $\varepsilon $
and the shape of the $A_{a}(t^{\prime })$.

\subsection{Orbits}

In this section, we investigate the contribution of individual trajectories
to the HHG cutoff of the homogeneous and nonhomogeneous cases by examining
the solutions of the saddle point equations. To get a better insight into
the problem, we employ a monochromatic field with $E(t)=E_{0}\sin (\omega
t)e_{x}$, where $e_{x}$ is the polarization vector along the $x$ axis. By
using the conventional relationship
\begin{equation}
\mathbf{A}(t)=-\int_{-\infty }^{t}E(t^{\prime })dt^{\prime }
\end{equation}%
and applying some trigonometric identities, the electric field (\ref%
{Newtonh1}) and potential field (\ref{newpotentialfield}) read
\begin{eqnarray}
E(t)&=&E_{0}\sin (\omega t)(1+2\alpha \sin (\omega t)/\omega ^{2}),
\label{monochelectricfield} \\
A_{eff}(t)&=&A_{0}\cos (\omega t)+2\alpha A_{a}(t),
\label{monochpostentialfield}
\end{eqnarray}%
respectively, where $A_{0}=E_{0}/\omega $ and
\begin{equation}
A_{a}(t)=A_{0}^{2}\sin (\omega t)/4\omega -A_{0}^{2}t/2.
\label{monochcorecpotentialfield}
\end{equation}%
In terms of the pondermotive energy of the homogeneous field, the drift
momentum of Eq. (\ref{driftmomentum}) yields
\begin{equation}
\mathbf{k}=-2\sqrt{U_{p}}\cos (\omega t^{\prime })+\epsilon\left[\frac{U_{p}}{%
\omega }\sin (\omega t^{\prime })-2U_{p}t^{\prime }\right](1\mp \sqrt{3})
\label{newdriftmomentum}
\end{equation}%
Based on the above equations, we solve the saddle point equations defined in
Eqs. (\ref{saddle1})- (\ref{saddle3}) in terms of the ionization ($t^{\prime}$) and recombination ($t$) times.

\begin{figure}[th]
\begin{center}
\includegraphics[width=0.4\textwidth]{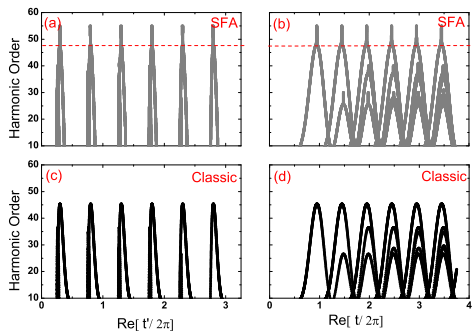}
\end{center}
\caption{(Color online) Dependence of harmonic order on the release time $%
t^{\prime }$ and the recombination time $t$ of the electron for the homogeneous field ($\protect%
\varepsilon=0$). We consider an hydrogen atom, for which the ground-state
energy is $I_{p}=0.5$ a.u., in a linearly polarized, monochromatic field of
frequency $\protect\omega=0.057$ a.u. and intensity $I=3\times10^{14}
\mathrm{W/cm^{2}}$. Panels (a) and (b) give the harmonic order as a function
of the ionization and recombination times of the SFA model, respectively,
while panels (c) and (d) depict the the harmonic order in terms of
ionization and recombination times of the classical calculations,
respectively. The red dashed lines correspond to the harmonic cutoff.}
\label{HHG1}
\end{figure}

\begin{figure}[th]
\begin{center}
\includegraphics[width=0.4\textwidth]{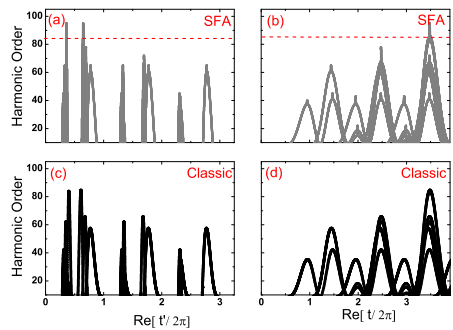}
\end{center}
\caption{(Color online) Dependence of harmonic order on the release time $t^{\prime }$ and the recombination time $t$ of the electron for a nonhomogeneous field with $\varepsilon=0.003$ using the
same parameters as in Fig.~\ref{HHG1}. Panels (a) and
(b) give the ionization and recombination times of SFA model, respectively,
while panels (c) and (d) depict the ionization and recombination times of
the classical calculations, respectively. The red dashed lines correspond
to the harmonic cutoff.}
\label{HHG4}
\end{figure}

In this paper, we present these solutions briefly, for more
detailed analysis we refer to our previous paper~\cite{tahirsfa}. In Figs. \ref{HHG1} and \ref{HHG4}, we plot the harmonic order as function
of the real parts of the ionization  ($t^{\prime }$) and recombination ($t$)
times for the case with $\varepsilon =0$ and $\varepsilon =0.003$ (panels a
and b for $t^{\prime }$ and $t$, respectively), respectively. In these
figures, in comparison to the SFA model, we also present the classical
solutions of $t^{\prime }$ and $t$ (panels c and d, respectively). From
these figures it is clear that the SFA resembles the classical
calculations. For the homogeneous case, the ionization and recombination
times corresponding to each cycle are identical and the largest cutoff (harmonic $48\omega $) comes from the shortest pairs.

For the nonhomogeneous case, the general harmonic cutoff is extended in
comparison to the homogeneous case, but the trajectories do not follow the
same symmetry as shown in Fig.~\ref{HHG1}.

\begin{figure}[h]
\begin{center}
\includegraphics[width=0.4\textwidth]{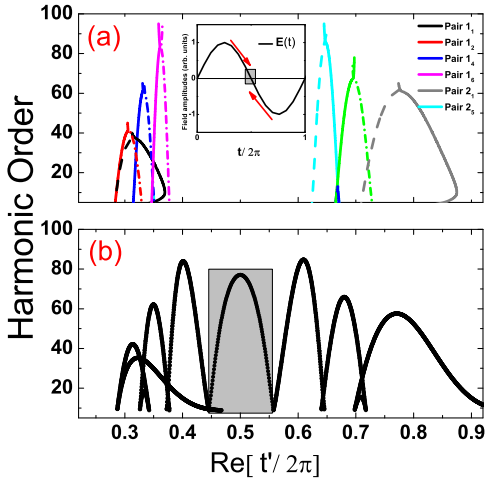}
\end{center}
\caption{(Color online) Harmonic order as a function of the release times $%
t^{\prime }$ of the electron for the same parameters given in Fig. \protect
\ref{HHG4}, but for pairs corresponding to the solutions from positive and negative sides of the first cycle. Panel (a) gives the SFA real part of the release times, while panel (b) depict the classical ionization times of
these pairs. The dashed and solid lines correspond to the long and the short
orbits, while the pairs with dot dashed lines do not have the well-known
shorts and long pairs.}
\label{HHG7}
\end{figure}

In here, the electron tunnels with two different canonical momenta given by
Eq. (\ref{newdriftmomentum}): one leads to a higher cutoff, and the other to
a lower one. For example, for the shortest pairs corresponding to the
electron leaving at the field maxima the cutoff is at around harmonic $38\omega$, while for the pairs corresponding to the electron leaving at the
field minima it is at around harmonic $60\omega $.

Furthermore, the cutoff become larger as we move from shorter pairs to the
longer pairs. As it demonstrated in Fig. \ref{HHG7} the trajectories from the minimum and maximum of the cycle moves towards
each other until they collocate on each other at the field crossing. As $\varepsilon$ increases the solutions
from positive and negative side of the cycle collapse faster to each other.
It means there will be less available pairs of orbit for electron to produce
HHG. In case of the field with very large $\varepsilon $ even the long
trajectory of the shortest orbits are not allowed as we demonstrated it using the actual enhanced field of the nanostructure~\cite{ciappipra}.

\subsection{Spectra}

In this section, we compute HHG spectra using Eq. (\ref{Mp}).
\begin{figure}[h]
\begin{center}
\includegraphics[width=0.4\textwidth]{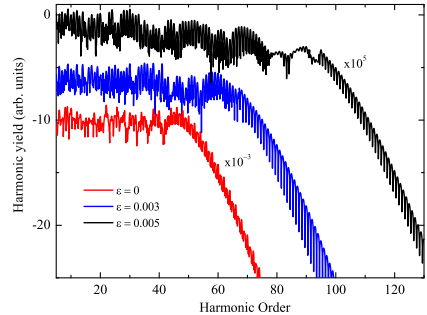}
\end{center}
\caption{(Color online) High-order harmonic spectra for hydrogen atoms ($%
I_{p}=0.5$ a.u.) interacting with a monochromatic field of frequency $%
\protect\omega=0.057$ a.u. and intensity $I=3\times10^{14} \mathrm{W/cm^{2}}$%
. Red, blue and black colors depict the cases with $\protect\varepsilon=0$, $%
\protect\varepsilon=0.003$ and $\protect\varepsilon=0.005$, respectively.
The HHG spectra obtained for $\protect\varepsilon=0$ and $\protect\varepsilon%
=0.005$ are scaled for clarity.}
\label{spectra3}
\end{figure}

These results are depicted in Fig.~\ref{spectra3} with red colored spectra presenting the homogeneous field and blue colored and black colored spectra showing the nonhomogeneous case with $\varepsilon =0.003$ and $\varepsilon =0.005$, respectively.
In all cases, the HHG cutoff is in good agreement with
the trajectories analysis represented in the previous section and with the
full 3D numerical calculations of Ref~\cite{yavuz}. These cutoffs are at around harmonics $45\omega $, $73\omega $ and $93\omega $ for
$\varepsilon =0$, $\varepsilon =0.003$ and $\varepsilon =0.005$, respectively. Moreover, it seems that both odd and
even harmonics are present in the total spectra. At higher regime of the spectra, both odd and even harmonic have the same weight, while at the lower regime one sets of harmonics are a bit more dominant than the others.

\section{Conclusions and outlook}

We present theoretical studies of high-order harmonic generation in
noble gases produced by spatially nonhomogeneous fields. These fields are generated when
different nanosystems, such as metal bow-tie nanostructures, metal and dielectric nanoparticles and
nanotips, are exposed to a short and intense laser pulse. In this study, we have employed
 a numerical approach as well as performed an exhaustive theoretical analysis.

For the numerical method, we used the numerical solution of the time dependent Schroedinger equation (TDSE) in reduced dimensions combined
with the actual functional form of the electric field extracted from finite element simulations to predict
the harmonic spectra originated when Ar atoms interact with the plasmonic enhanced field in a metal bow-tie nanostructure. 
We observe an extension in the harmonic cutoff position that could lead to the production of XUV coherent laser sources and opening the avenue to the
generation of attosecond pulses. For this particular nanostructure, this new feature is a consequence of the combination of a
nonhomogeneous electric field, which modifies substantially the electron trajectories,
and the confinement of the electron dynamics. 

In our theoretical analysis, we employed both classical and semiclassical approaches to investigate the high-order harmonic
spectra produced by spatially nonhomogeneous fields of linear form. We show how the quantum orbits manifest themselves in these
particular spatially inhomogeneous fields. In addition we predict that in nonhomogeneous fields, the
electron tunnels with two different canonical momenta leading to a higher and lower cutoff. We also conclude that in the case of linear nonhomogeneous fields, both odd and even harmonics are present in the HHG spectra. Within our model, we show that the HHG cutoff
extends to the larger harmonics as a function of the inhomogeneity strength and this feature is consistent with the findings recognized
using actual fields.

\section*{Acknowledgments}

We acknowledge the financial support of the MICINN project FIS2008-00784
TOQATA; ERC Advanced Grant QUAGATUA, Alexander von Humboldt Foundation and
Hamburg Theory Prize (M. L.). This research has been partially supported by
Fundaci\'o Privada Cellex. We thanks Peter Hommelhoff and Samuel Markson for useful comments
and suggestions. 


\end{document}